\documentclass[prb,twocolumn,showpacs]{revtex4}
\usepackage{amsmath,amsthm,amssymb}
\usepackage{latexsym}
\usepackage{array}
\usepackage{amscd,graphicx}
\usepackage{bm,textcomp}

\begin{document}

\title{Influences of spin accumulation on the intrinsic spin Hall effect in
two dimensional electron gases with Rashba spin-orbit coupling}
\author{Xiaohua Ma$^{1,2}$, Liangbin Hu$^{2,3}$ , Ruibao Tao$^{1}$, and
Shun-Qing Shen$^{1,2}$} \affiliation{$^{1}$State Key Laboratory of
Applied Surface Physics and Department of Physics, Fudan University, Shanghai 200433, China \\
$^{2}$Department of Physics, The University of Hong Kong, Pokfulam
Road,
Hong Kong, China \\
$^{3}$Department of Physics, South China Normal University, Guangdong
510631, China }
\date{\today }

\begin{abstract}
In a two dimensional electron gas with Rashba spin-orbit coupling, the
external electric field may cause a spin Hall current in the direction
perpendicular to the electric field. This effect was called the intrinsic
spin Hall effect. In this paper, we investigate the influences of spin
accumulation on this intrinsic spin Hall effect. We show that due to the
existence of boundaries in a real sample, the spin Hall current generated by
the intrinsic spin Hall effect will cause spin accumulation near the edges
of the sample, and in the presence of spin accumulation, the spin Hall
conductivity will not have a universal value. The influences of spin
accumulation on the intrinsic spin Hall effect in narrow strips of two
dimensional electron gases with Rashba spin-orbit coupling are investigated
in detail.
\end{abstract}

\maketitle


\section{Introduction}

Spintronics, which aims at the manipulation of the electron's spin degree of
freedom in electronic devices, has become an emerging field of condensed
matter physics for its potential application in information industry.\cite%
{ga,sdr,Ball} Though originally many spintronic concepts involve
ferromagnetic metals since spins in this kind of system behave collectively
and hence are easier to be controlled,\cite{mb,jltr,mr,johnson,ta}
spintronics in semiconductors is more interesting because dope and
heterojunction formation can be used to obtain some specific devices. It is
anticipated that combining the advantages of semiconductors with the
concepts of spintronics will yield fascinating new electronic devices and
open the way to a new field of physics, i.e., semiconductor spintronics.
However, at present many great challenges still remain in this exciting
quest. Among them, an issue that is fundamentally important in semiconductor
spintronics and has not been resolved is how to achieve efficient injections
of spins into non-magnetic semiconductors at room temperature\cite%
{Ham99,Schmd,Ohno,Fdl,Mat03}. Usage of ferromagnetic metals as sources of
spin-injection is not practical because most of the spin-polarizations will
be lost at the interface between metal and semiconductor due to the large
conductivity mismatch\cite{Ham99,Schmd}. Another possible approach is to use
ferromagnetic semiconductors ( such as Ga$_{1-x}$Mn$_x$As ) instead of
ferromagnetic metals as sources of spin-injection. In this approach, the
problem of conductivity mismatch does not exist and hence efficient
injections of spins into non-magnetic semiconductors can truly be achieved
\cite{Ohno,Fdl,Mat03}. But for practical use at room temperature, the Curie
temperatures of ferromagnetic semiconductors are still too low. Thus, from
both the experimental and theoretical points of view, more great efforts are
still needed in order to achieve efficient injections of spins in
non-magnetic semiconductors at room temperature. Recently, a surprising
effect was predicted theoretically that an electric field may cause a
quantum spin Hall current in the direction perpendicular to the electric
field in conventional hole-doped semiconductors ( such as Si, Ge, and GaAs )
\cite{sns} or in two dimensional electron gases ( 2DEGs ) with Rashba
spin-orbit coupling.\cite{jdq} This intrinsic $spin$ Hall effect might offer
a new approach for achieving efficient injections of spins in non-magnetic
semiconductors and reveal a new avenue in the spintronics research.

In this paper, we study the influences of spin accumulation on the intrinsic
spin Hall effect in two dimensional electron gases with Rashba spin-orbit
coupling. From the standpoint of spintronic applications, it is important to
understand whether the spin Hall currents predicted in Refs.[14-15] are $%
transport$ spin currents, i.e., whether they can be employed for
transporting spins. An important feature of transport spin currents is that
they will induce nonequilibrium spin accumulation at some specific
locations, for example, at the boundaries of a sample or at the interfaces
between two different materials\cite{mr,johnson,ta}. On the other hand, if
spin accumulation was caused in a sample due to the flow of a spin current,
the spin current will also be changed significantly by the spin accumulation.%
\cite{mr,johnson,ta} So in a strict theoretically treatment of the intrinsic
spin Hall effect, the interplay between the spin Hall current and the spin
accumulation must be taken into account. A detailed theoretical
investigation of the influences of spin accumulation on the intrinsic spin
Hall effect in hole-doped semiconductors was presented in Ref.[16]. In the
present paper, we will use a similar method as was applied in Ref.[16] to
investigate the influences of spin accumulation on the intrinsic spin Hall
effect in two dimensional electron gases with Rashba spin-orbit coupling. We
will show that in contrast with what was found in Refs.[15, 17-18], in the
presence of spin accumulation, the spin Hall conductivity in the intrinsic
spin Hall effect in a Rashba two dimensional electron gas $does$ $not$ have
a universal value, and in order to calculate correctly the spin Hall current
and the spin Hall conductivity in a real sample with boundaries, the
influences of spin accumulation need to be taken into account. The paper
will be organized as follows: In Sec.II, we will first present a brief
introduction to the intrinsic spin Hall effect in a Rashba two dimensional
electron gas in the absence of spin accumulation and impurity scattering. In
Sec.III, the influences of spin accumulation and impurity scattering will be
taken into account. In Sec.IV, by use of the formulas derived in Sec.III,
the influences of spin accumulation on the intrinsic spin Hall effect in a
narrow strip of a Rashba two dimensional electron gas will be discussed in
detail.

\section{Intrinsic spin Hall effect in the absence of spin accumulation and
impurity scattering}

In this paper, we will use a slightly different method from what was applied
in Refs.[15,17-18] to discuss the intrinsic spin Hall effect in two
dimensional electron gases with Rashba spin-orbit coupling. The merit of
this method is that the influences of spin accumulation can be easily
included. For clarity, in this section we first present a brief introduction
to the intrinsic spin Hall effect in the absence of spin accumulation and
impurity scattering. We will show that our method will yield the same
results as was obtained in Refs.[15,17-18] if spin accumulation and impurity
scattering are neglected.

In the momentum representation, the single-particle Hamiltonian for a two
dimensional electron gas with Rashba spin-orbit coupling reads\cite{Rash},
\begin{eqnarray}
\hat{H}_0 &=&\frac{\hbar ^2k^2}{2m}-\alpha (\mathbf{z}\times \mathbf{k}%
)\cdot \mathbf{\sigma }  \notag \\
&=&\frac{\hbar ^2k^2}{2m}-\alpha k(\cos \theta \sigma _y-\sin \theta \sigma
_x),
\end{eqnarray}
where $\mathbf{z}$ is the unit vector along the normal of the
two-dimensional plane, $\mathbf{k}=(k\cos \theta ,k\sin \theta )$ is the
momentum of an electron, $\mathbf{\sigma =}(\sigma _x,\sigma _y,\sigma _z)$
the Pauli matrices, and $\alpha $ the Rashba spin-orbit coupling constant.
The Hamiltonian (1) can be diagonalized exactly. The eigenenergies are given
by
\begin{equation}
\epsilon _{\mathbf{k}\lambda }=\frac{\hbar ^2k^2}{2m}-\lambda \alpha
k,\qquad (\lambda =\pm 1),
\end{equation}
and the corresponding spinor eigenstates are defined by
\begin{equation}
|\mathbf{k}\lambda \rangle =\frac 1{\sqrt{2}}%
\begin{pmatrix}
-i \lambda e^{-i\theta } \\
1%
\end{pmatrix}%
.\qquad
\end{equation}
The expectation value of the spin of an electron which is in the spinor
eigenstate $|\mathbf{k}\lambda \rangle $ will be given by
\begin{eqnarray}
\mathbf{S}_\lambda ^{(0)}(\mathbf{k}) &=&\frac \hbar 2\langle \mathbf{k}%
\lambda |\mathbf{\sigma }|\mathbf{k}\lambda \rangle  \notag \\
&=&S_{\lambda ,x}^{(0)}(\mathbf{k})\mathbf{e}_x+S_{\lambda ,y}^{(0)}(\mathbf{%
k})\mathbf{e}_y,
\end{eqnarray}
where $\mathbf{e}_x$ and $\mathbf{e}_y$ are the unit vectors along the $x$
and $y$ axes, respectively, and $S_{\lambda ,x}^{(0)}(\mathbf{k})$ and $%
S_{\lambda ,y}^{(0)}(\mathbf{k})$ are the $x$ and $y$ components of the
spin, which are given by
\begin{eqnarray}
S_{\lambda ,x}^{(0)}(\mathbf{k}) &=&\frac \hbar 2\langle \mathbf{k}\lambda |%
\hat{\sigma}_x|\mathbf{k}\lambda \rangle =\lambda \frac{\hbar k_y}{2k},\qquad
\\
S_{\lambda ,y}^{(0)}(\mathbf{k}) &=&\frac \hbar 2\langle \mathbf{k}\lambda |%
\hat{\sigma}_y|\mathbf{k}\lambda \rangle =-\lambda \frac{\hbar k_x}{2k}.
\end{eqnarray}
Eqs.(4-6) show that the eigenstates of the Hamiltonian (1) are
spin-polarized in the in-plane directions, and the spin-polarization
directions depend on the momentum $\mathbf{k}$.

When an external electric field is applied in the $x$ direction, the total
Hamiltonian of the system will be given by $\hat{H}=\hat{H}_{0}+e\mathbf{E}%
\cdot \mathbf{x}$, where $\mathbf{E=}E_{x}\mathbf{e}_{x}$ is the external
electric field and $-e$ the charge of the electron. The equation of motion
for the electron's position and spin degrees of freedom under the action of
the external electric field can be obtained directly from the Heisenberg
equation, and the results read
\begin{eqnarray}
\frac{d}{dt}\mathbf{k} &=&\frac{e\mathbf{E}}{\hbar }, \\
\frac{d}{dt}\mathbf{S} &=&\mathbf{h}_{eff}\times \mathbf{S,}
\end{eqnarray}%
where $\mathbf{h}_{eff}=\frac{2\alpha }{\hbar }(k_{y}\mathbf{e}_{x}-k_{x}%
\mathbf{e}_{y})$ is an effective magnetic field felt by an electron with
momentum $\mathbf{k}$ due to the Rashba spin-orbit coupling. The time
variation of the spin of an electron which is initially in the spinor
eigenstate $|\mathbf{k\lambda \rangle }$ ( i.e., $\mathbf{S(\mathit{t=0})=}%
\frac{\hbar }{2}\mathbf{\langle k\lambda }|\mathbf{\hat{\sigma}|k\lambda
\rangle }$ ) can be got by solving Eqs.(7-8) simultaneously. In this paper
we will consider only the linear response of the transport property to the
electric field. \ In the linear response regime, Eqs.(7-8) can be integrated
analytically by use of the same method of Ref.[15], and one can find that
under the action of the external electric field, the spin of an electron
with momentum $\mathbf{k}$ and initially in the spinor eigenstate $|\mathbf{%
k\lambda \rangle }$ will become
\begin{eqnarray}
S_{\lambda ,x}(\mathbf{k}) &\simeq &S_{\lambda ,x}^{(0)}(\mathbf{k)}, \\
S_{\lambda ,y}(\mathbf{k}) &\simeq &S_{\lambda ,y}^{(0)}(\mathbf{k}), \\
S_{\lambda ,z}(\mathbf{k}) &\simeq &-\lambda \frac{e\hbar k_{y}}{4\alpha
k^{3}}E_{x},
\end{eqnarray}%
where $S_{\lambda ,i}(\mathbf{k})$ is the $i$th component of the spin.
Eqs.(9-11) show that an applied electric field in the $x$ direction will
cause the spin to tilt in the perpendicular direction by an amount
proportional to $k_{y}$. Due to this fact, the application of an external
electric field in the $x$ direction will induce a spin Hall current in the $%
y $ direction with spin parallel to the $z$ direction. The spin Hall current
can be calculated by the following formula
\begin{equation}
J_{y}^{S_{z}}=\sum_{\lambda }\int \frac{d^{2}\mathbf{k}}{(2\pi )^{2}}%
S_{\lambda ,z}(\mathbf{k})(\hbar k_{y}/m)f_{\lambda }(\mathbf{k}),
\end{equation}%
where $f_{\lambda }(\mathbf{k})$ is the probability distribution function of
conduction electrons. If spin accumulation and electron-impurity scattering
can be neglected, $f_{\lambda }(\mathbf{k})$ can be given simply by the
Fermi-Dirac equilibrium distribution function, i.e., $f_{\lambda }(\mathbf{k}%
)=f^{0}(\epsilon _{\mathbf{k}\lambda })\equiv 1/\{\exp [\beta (\epsilon _{%
\mathbf{k}\lambda }-\epsilon _{F})]+1\}$, where $\beta =1/k_{B}T$ and $%
\epsilon _{F}$ is the Fermi energy. Then one can find that in the usual case
where both spin-orbit split bands are occupied, the spin Hall current $%
J_{y}^{S_{z}}$ and the spin Hall conductivity $\sigma _{s}$ will be given by%
\cite{jdq}
\begin{equation}
J_{y}^{S_{z}}=\sigma _{s}E_{x},\;\sigma _{s}=\frac{e}{8\pi }.
\end{equation}%
Eq.(13) is the central result of Ref.[15], which was also obtained by
several other groups with different theoretical approaches.\cite{Others}
Eq.(13) shows that as long as both spin-orbit split bands are occupied, the
spin Hall conductivity will have a $universal$ value, independent of both
the Rashba spin-orbit coupling strength and of the density of conduction
electrons. Though Eq.(13) was obtained in the clean limit, recent numerical
simulation shows that it still holds in the presence of weak disorder,
providing that the sample size exceeds the localization length.\cite{xiexc}
It is also important to note that unlike the similar effect conceived by
Hirsch\cite{hirsch}, which is caused by spin-orbit dependent anisotropic
scattering from impurities and will vanish in the weak scattering limit,\cite%
{hirsch, zhang, Hu} the spin Hall effect described by Eq.(13) has a quantum
nature and is purely intrinsic, i.e., it does not rely on any anisotropic
scattering from impurities. Of course, it should be pointed out that though
the mechanism of the intrinsic spin Hall effect described above does not
involve impurity scattering, it does not mean that impurity scattering have
no significant influences on the effect. The reason is that in a real
sample, due to the existence of boundaries, nonequilibrium spin accumulation
will be caused inevitably near the edges of the sample when the spin Hall
current circulates in it, and in the presence of nonequilibrium spin
accumulation, spin diffusion will be induced by electron-impurity scattering
and, hence, the spin Hall current may also be changed significantly from
what was given by Eq.(13). Thus in order to calculate correctly the spin
Hall current and the spin Hall conductivity in a real sample with
boundaries, the influences of spin accumulation and electron-impurity
scattering need to be considered.

\section{Influences of spin accumulation and impurity scattering}

In the presence of spin accumulation and electron-impurity scattering, the
distribution function of conduction electrons can no longer be given simply
by the Fermi-Dirac equilibrium distribution function but should be derived
strictly by solving the Boltzmann transport equation. In a steady ( but
nonequilibrium ) state, the Boltzmann equation reads
\begin{align}
\mathbf{V}_\lambda (\mathbf{k})& \cdot \mathbf{\nabla }f_\lambda (\mathbf{r},%
\mathbf{k})-e\mathbf{E}_{ext}\cdot \mathbf{V}_\lambda (\mathbf{k})\frac{%
\partial f_\lambda (\mathbf{r},\mathbf{k})}{\partial \epsilon _\lambda (k)}
\notag \\
& =-[(\frac{\partial f_\lambda }{\partial t})_{coll.}^{(\lambda \rightarrow
\lambda )}+(\frac{\partial f_\lambda }{\partial t})_{coll.}^{(\lambda
\rightarrow \bar{\lambda})}],  \label{Eq:be}
\end{align}
where $\mathbf{V}_\lambda (\mathbf{k})=\frac 1\hbar \mathbf{\nabla }_{%
\mathbf{k}}\epsilon _{\mathbf{k}\lambda }$ is the velocity of conduction
electrons, $\mathbf{E}_{ext}=E_x\mathbf{e}_x$is the external electric field
applied in the $x$ direction, and $f_\lambda (\mathbf{r},\mathbf{k})$ is the
distribution function. The collision term $(\partial f_\lambda /\partial
t)_{coll.}^{(\lambda \rightarrow \lambda ^{^{\prime }})}$ describes the
changes of the distribution function due to the intra-band ( $\lambda
^{^{\prime }}=\lambda $ ) and/or inter-band ( $\lambda ^{^{\prime }}=\bar{%
\lambda}$ ) electron-impurity scattering, which is given by
\begin{align}
(\frac{\partial f_\lambda }{\partial t})_{coll}^{(\lambda \rightarrow
\lambda ^{^{\prime }})}=& -\int \frac{d^2k^{\prime }}{(2\pi )^2}\omega
_{\lambda ,\lambda ^{^{\prime }}}(\mathbf{k},\mathbf{k^{\prime }})\delta
(\epsilon _\lambda (\mathbf{k})-\epsilon _{\lambda ^{^{\prime }}}(\mathbf{%
k^{\prime }}))  \notag \\
& \times [f_\lambda (\mathbf{r},\mathbf{k})-f_{\lambda ^{^{\prime }}}(%
\mathbf{r},\mathbf{k^{\prime }})],  \label{Eq:coll2}
\end{align}
where $\omega _{\lambda ,\lambda ^{^{\prime }}}(\mathbf{k},\mathbf{k^{\prime
}})$ is the rates of an electron to be scattered from the state $|\mathbf{k}%
\lambda \rangle $ into the state $|\mathbf{k^{\prime }}\lambda ^{^{\prime
}}\rangle $ by impurity scattering.

The Boltzmann equation (14) can be solved by the relaxation time
approximation method. Within the relaxation time approximation and in the
linear response regime, the system can be considered only slightly deviated
from the equilibrium state, thus the total distribution $f_{\lambda }(%
\mathbf{r},\mathbf{k})$ can be expressed as the sum of the equilibrium
distribution function $f^{0}(\epsilon _{\mathbf{k}\lambda })$ and the
non-equilibrium ones as the following
\begin{align}
f_{\lambda }(\mathbf{r},\mathbf{k})=& f^{0}(\epsilon _{\mathbf{k}\lambda
})-e\mu ^{\lambda }(\mathbf{r})\frac{\partial f^{0}(\epsilon _{\mathbf{k}%
\lambda })}{\partial \epsilon _{\mathbf{k}\lambda }}  \notag \\
& +e\tau _{\lambda }(\mathbf{k})\mathbf{E}^{\lambda }(\mathbf{r})\cdot
\mathbf{V}_{\lambda }(\mathbf{k})\frac{\partial f^{0}(\epsilon _{\mathbf{k}%
\lambda })}{\partial \epsilon _{\mathbf{k}\lambda }}.  \label{Eq:f}
\end{align}%
Here the second term denotes the change of the distribution function due to
the occurrence of nonequilibrium spin accumulation in the sample, with $%
-e\mu ^{\lambda }(\mathbf{r})$ ( $\lambda =\pm $ ) denoting the band- and
position-dependent shifts of the Fermi level in the nonequilibrium state,
which characterize the imbalance of the filling of conduction electrons in
the two spin-orbit split bands in the presence of nonequilibrium spin
accumulation. The third term in Eq.(16) denotes the changes of the
distribution function due to the movement of conduction electrons under the
action of an effective electric field $\mathbf{E}^{\lambda }(\mathbf{r})$,
with $\tau _{\lambda }(\mathbf{k})$ denoting the total relaxation time of
conduction electrons with momentum $\mathbf{k}$, which is determined by the
electron-impurity scattering. Because the occurrence of nonequilibrium spin
accumulation will cause spin diffusion in the sample, the effective electric
field $\mathbf{E}^{\lambda }(\mathbf{r})$ should be given by $\mathbf{E}%
^{\lambda }(\mathbf{r})=\mathbf{E}_{ext}-\mathbf{\nabla }\mu ^{\lambda }(%
\mathbf{r})$, i.e., in addition to the external electric field $\mathbf{E}%
_{ext}$, conduction electrons will also feel an effective field given by the
gradients of the band- and position-dependent shifts of the Fermi level.\cite%
{mr,johnson,ta} Substitute Eq.(\ref{Eq:f}) into Eqs.(14-15), the left-hand
side of the Boltzmann equation will become
\begin{eqnarray}
Lhs &=&-e\frac{\partial f^{0}(\epsilon _{\mathbf{k}\lambda })}{\partial
\epsilon _{\mathbf{k}\lambda }}\{\mathbf{E}^{\lambda }(\mathbf{r})\cdot
\mathbf{V}_{\lambda }(\mathbf{k})  \notag \\
&&-\tau _{\lambda }(\mathbf{k})\mathbf{V}_{\lambda }(\mathbf{k})\cdot
\bigtriangledown \lbrack \mathbf{E}^{\lambda }(\mathbf{r})\cdot \mathbf{V}%
_{\lambda }(\mathbf{k})]\},
\end{eqnarray}%
and the right-hand side of the Boltzmann equation become
\begin{align}
Rhs& =-e\frac{\partial f^{0}(\epsilon _{\mathbf{k}\lambda })}{\partial
\epsilon _{\mathbf{k}\lambda }}\{\tau _{\lambda }(\mathbf{k})[\frac{1}{\tau
_{\lambda }^{\uparrow \uparrow }(\mathbf{k})}+\frac{1}{\tau _{\lambda
}^{\uparrow \downarrow }(\mathbf{k})}][\mathbf{E}^{\lambda }(\mathbf{r}%
)\cdot \mathbf{V}_{\lambda }(\mathbf{k})]  \notag \\
& -\frac{\mu ^{\lambda }(\mathbf{r})-\mu ^{-\lambda }(\mathbf{r})}{\tau
_{\lambda }^{\uparrow \downarrow }(\mathbf{k})}\},
\end{align}%
where $\tau _{\lambda }^{\uparrow \uparrow }(\mathbf{k})$ and $\tau
_{\lambda }^{\uparrow \downarrow }(\mathbf{k})$ are the intra-band and
inter-band transition relaxation times, respectively, which are defined by
\begin{align}
& \tau _{\lambda }^{\uparrow \uparrow }(\mathbf{k})=[\int \frac{%
d^{2}k^{\prime }}{(2\pi )^{2}}\omega _{\lambda ,\lambda }(\mathbf{k},\mathbf{%
k^{\prime }})\delta (\epsilon _{\lambda }(\mathbf{k})-\epsilon _{\lambda }(%
\mathbf{k^{\prime }}))]^{-1} \\
& \tau _{\lambda }^{\uparrow \downarrow }(\mathbf{k})=[\int \frac{%
d^{2}k^{\prime }}{(2\pi )^{2}}\omega _{\lambda ,-\lambda }(\mathbf{k},%
\mathbf{k^{\prime }})\delta (\epsilon _{\lambda }(\mathbf{k})-\epsilon
_{-\lambda }(\mathbf{k^{\prime }}))]^{-1}.
\end{align}%
By integrating both the left-hand side and the right-hand side of the
Boltzmann equation, one can find that the total relaxation time $\tau
_{\lambda \text{ }}$of conduction electrons should be given by
\begin{equation}
\tau _{\lambda }(\mathbf{k})=[\frac{1}{\tau _{\lambda }^{\uparrow \uparrow }(%
\mathbf{k})}+\frac{1}{\tau _{\lambda }^{\uparrow \downarrow }(\mathbf{k})}%
]^{-1},
\end{equation}%
and the band- and position-dependent shifts of the Fermi level satisfy the
following equation,
\begin{equation}
\bigtriangledown ^{2}\mu ^{\lambda }(\mathbf{r})=\frac{\mu ^{\lambda }(%
\mathbf{r})-\mu ^{-\lambda }(\mathbf{r})}{D_{\lambda }^{2}},  \label{Eq:diff}
\end{equation}%
where $D_{\lambda }\equiv \lbrack (V_{F}^{\lambda })^{2}\tau _{F}\tau
_{F}^{\uparrow \downarrow }]^{1/2}$, with $V_{F}^{\lambda }$ denoting the
band-dependent Fermi velocity and $\tau _{F}^{\uparrow \downarrow }$ the
inter-band-transition relaxation time and $\tau _{F}$ the total relaxation
time of conduction electrons at the Fermi level, respectively. ( For
simplicity, we assume that $\tau _{F}$ and $\tau _{F}^{\uparrow \downarrow }$
are band-independent. ). From Eq.(22), one can see that the relative shifts
of the Fermi level in the two spin-orbit split bands, given by $\mu ^{+}(%
\mathbf{r})-\mu ^{-}(\mathbf{r})$, satisfy the following diffusion equation,
\begin{equation}
\bigtriangledown ^{2}[\mu ^{+}(\mathbf{r})-\mu ^{-}(\mathbf{r})]=\frac{\mu
^{+}(\mathbf{r})-\mu ^{-}(\mathbf{r})}{D^{2}},
\end{equation}%
where $D$ is the spin-diffusion length, defined by
\begin{equation}
D=[\frac{1}{D_{+}^{2}}+\frac{1}{D_{-}^{2}}]^{-1/2}.
\end{equation}%
In addition to Eqs.(22-23), the band-dependent shifts of the Fermi level
should also satisfy the charge neutrality condition, which requires that the
net changes of the charge density due to the band-dependent shifts of the
Fermi level, given by $\delta \rho (\mathbf{r})=e\sum_{\lambda }\int \frac{%
d^{2}k}{(2\pi )^{2}}[f_{\lambda }(\mathbf{r},\mathbf{k})-f_{\lambda
}^{0}(\epsilon _{\mathbf{k\lambda }})]$, should be zero. This requirement
arises from the fact that according to the symmetry of the Hamiltonian (1),
in the direction perpendicular to the external electric field, no charge
Hall current will be generated, so the occurrence of spin accumulation due
to the flow of the spin Hall current does not result in charge accumulation.
Due to this requirement, one can show that in addition to Eqs.(22-23), the
band-dependent shifts of the Fermi level should also satisfy the following
equation,
\begin{equation}
\sum_{\lambda =\pm }k_{\lambda F}\mu ^{\lambda }(\mathbf{r})=0,
\end{equation}%
where $k_{\lambda F}$ is the band-dependent wave number at the Fermi level.
After the band-dependent shifts of the Fermi level are determined, the spin
Hall current can be obtained by inserting Eq.(11) and Eq.(16) into Eq.(12),
then one can find that in the usual case where both spin-orbit split bands
are occupied\cite{note1}, the spin Hall current and the spin Hall
conductivity will be given by

\begin{eqnarray}
J_y^{S_z}{}(\mathbf{r}) &=&\sigma _s(\mathbf{r})E_x, \\
\sigma _s(\mathbf{r}) &=&\frac e{8\pi }-\frac{e^2[\mu ^{+}(\mathbf{r})-\mu
^{-}(\mathbf{r)]}}{16\pi \alpha [(m\alpha /\hbar ^2)^2+2m\epsilon _F/\hbar
^2]^{1/2}}\mathbf{.}
\end{eqnarray}
Eqs.(26-27) show that the spin accumulation may have some significant
influences on the intrinsic spin Hall effect in a Rashba two dimensional
electron gas. Firstly, in the presence of spin accumulation, the spin Hall
conductivity will be a $position-dependent$ quantity and $do$ $not$ have a
universal value, i.e., it will depend on the Rashba spin-orbit coupling
constant and on the density of conduction electrons. This is very different
from what was shown in Eq.(13). Secondly, in the presence of spin
accumulation, the spin Hall current may be decreased substantially from the
corresponding value obtained in the absence of spin accumulation, and the
decrease will be determined by $\mu ^{+}(\mathbf{r})-\mu ^{-}(\mathbf{r)}$,
i.e., proportional to the relative shifts of the Fermi level in the two
spin-orbit split bands.

\section{Results and Discussions}

Eqs.(23,25) and (26--27) constitute a set of self-consistent equations, from
which both the spin Hall current and the spin accumulation can be obtained.
In this section, we apply these formulas to discuss the intrinsic spin Hall
in a narrow strip of a two dimensional electron gas with Rashba spin-orbit
coupling. Narrow strips are the usual geometry applied in the experimental
measurement of the Hall effect, including the spin Hall effect.\cite{hirsch,
zhang, Hu} In the following we will assume that the longitudinal direction
of the strip is along the $x$ axis and the transverse direction along the $y$
axis and the normal of the 2D plane along the $z$ axis, respectively, and an
external electric field $E_{x}$ is applied in the longitudinal direction of
the strip. According to Eqs.(26)-(27), in order to calculate the spin Hall
current $J_{y}^{S_{z}}$ caused by the longitudinal external electric field $%
E_{x}$, one must first find out the band-dependent shifts $\mu ^{+}(\mathbf{r%
})$ and $\mu ^{-}(\mathbf{r)}$ of the Fermi level. For simplicity, we assume
that the length $L$ of the strip is much larger than its width $w$ so that
spin diffusion in the longitudinal direction of the strip can be neglected.
In such case, only transverse spin accumulation need to be considered, and $%
\mu ^{\lambda }(\mathbf{r})$, $J_{y}^{S_{z}}{}(\mathbf{r})$, and $\sigma
_{s}(\mathbf{r})$ will all depend only on the $y$ coordinates. From Eq.(23),
$\mu ^{+}(y)-\mu ^{-}(y)$ can be expressed as
\begin{equation}
\mu ^{\lambda }(y)-\mu ^{-\lambda }(y)=Ae^{y/D_{1}}+Be^{-y/D_{1}},
\label{Eq:s1}
\end{equation}%
where $A$ and $B$ are two constant coefficients that need to be determined
by appropriate boundary conditions. In this paper, we will consider the
transverse-open-circuit boundary condition. In the transverse-open-circuit
boundary condition, the spin Hall current at the two boundaries of the
sample, which are assumed to be located at $y=\pm \frac{w}{2}$, should be
zero, i.e.,
\begin{equation}
J_{y}^{S_{z}}(y=\pm \frac{w}{2})=0.  \label{Eq:bc}
\end{equation}%
Substituting Eq.(\ref{Eq:s1}) into Eqs.(26-27) and by use of the above
boundary condition, the coefficients $A$ and $B$ can be determined. Then the
band- and position-dependent shifts of the Fermi level in the strip can be
obtained, and we get
\begin{align}
& \mu ^{+}(y)=\frac{M_{0}\cosh (y/D)}{2\cosh (w/2D)}[1+\frac{D^{2}}{D_{0}^{2}%
}], \\
& \mu ^{-}(y)=\frac{M_{0}\cosh (y/D)}{2\cosh (w/2D)}[\frac{D^{2}}{D_{0}^{2}}%
-1],
\end{align}%
where $M_{0}$ and $D_{0}$ are defined by
\begin{eqnarray}
M_{0} &=&\frac{2\alpha }{e}\sqrt{(m\alpha /\hbar ^{2})^{2}+2m\epsilon
_{F}/\hbar ^{2}}, \\
D_{0} &=&[\frac{1}{D_{+}^{2}}-\frac{1}{D_{-}^{2}}]^{-1/2}.
\end{eqnarray}%
Here $D_{\lambda }$ ( $\lambda =\pm $ ) has been defined in Eq.(22). In
obtaining Eqs.(30-31), the charge neutrality condition (25) was also used.
After $\mu ^{\lambda }(y)$ is determined, according to Eqs.(26-27), the spin
Hall current and the spin Hall conductivity will also be obtained, and the
results read
\begin{eqnarray}
J_{y}^{S_{z}}(y) &=&\sigma _{s}(y)E_{x,} \\
\sigma _{s}(y) &=&\frac{e}{8\pi }[1-\frac{\cosh (y/D)}{\cosh (w/2D)}].
\end{eqnarray}%
Eqs.(34-35) show that, due to the influences of spin accumulation, the
spatial distribution of the spin Hall current in a sample will be highly
inhomogeneous and the spin Hall conductivity is sensitively
position-dependent. The spin Hall current and the spin Hall conductivity and
their spatial distributions will also have sensitive dependences on the spin
diffusion length $D$ and the sample width $w$. This was shown in Fig.1,
where we have plotted the transverse spatial distributions of the spin Hall
currents in three distinct cases with different ratios of $D/w$. From Fig.1,
one can see that if $w\ll D$, the spin Hall current will be negligibly small
in the sample, i.e., $J_{y}^{S_{z}}(y)\simeq 0$ everywhere. On the other
hand, if $w\gg D$, the spin Hall conductivity will be approximately a
constant at $|y|\ll w/2$, i.e., $\sigma _{s}(y)=J_{y}^{S_{z}}(y)/E_{x}\simeq
$ $\frac{e}{8\pi }$ at $|y|\ll w/2$, which is independent of both the Rashba
spin-orbit coupling strength and of the density of conduction electrons. But
$\sigma _{s}(y)$ will decrease to zero as $y\rightarrow \pm w/2$.

The spin accumulation caused by the longitudinal electric field $E_x$ can be
calculated through the following formula
\begin{equation}
\langle \mathbf{S}\rangle =\sum_\lambda \int \frac{d^2k}{(2\pi )^2}\mathbf{S}%
_\lambda (\mathbf{k})f_\lambda (\mathbf{r},\mathbf{k}),
\end{equation}
where $\mathbf{S}_\lambda (\mathbf{k})$ has been given in Eqs.(9-11). By
inserting Eq.(16) and Eqs.(9-11) into Eq.(36) and with the help of
Eqs.(30-31), one can find that both the $y$ component of $\langle \mathbf{S}%
\rangle $ ( the in-plane spin accumulation ) and the $z$ component of $%
\langle \mathbf{S}\rangle $ ( the perpendicular spin accumulation ) are
non-zero,
\begin{eqnarray}
\langle S_y\rangle &=&\frac{em\alpha \tau _FE_x}{4\pi \hbar ^2}, \\
\langle S_z\rangle &=&\frac{eE_x}{4\pi }(\frac{m\tau _F}{\tau _F^{\uparrow
\downarrow }\epsilon _F})^{1/2}(1+\frac{\alpha ^2m}{\hbar ^2\epsilon _F})%
\frac{\sinh (y/D)}{\cosh (w/2D)}.
\end{eqnarray}
Eqs.(37-38) show that both the in-plane and the perpendicular spin
accumulation are proportional to the longitudinal electric field $E_x$, but
there are some significant differences between them. The in-plane spin
accumulation is homogeneously distributed in the sample and independent of
both the spin diffusion length $D$ and of the sample width $w$. However, the
spatial distribution of the perpendicular spin accumulation is highly
inhomogeneous and its magnitude depends sensitively on the spin diffusion
length $D$ and on the sample width $w$. These differences arise from the
fact that the in-plane and the perpendicular spin accumulation are caused by
very different mechanism. In fact, it was known long time ago that in a two
dimensional electron gas with Rashba spin-orbit coupling, an applied
in-plane electric field will induce a homogeneous in-plane spin accumulation
polarized in the direction perpendicular to the electric field\cite{Edel,
Inoue}, but the in-plane spin accumulation has nothing to do with the
intrinsic spin Hall effect. From the theoretical viewpoints, the in-plane
spin accumulation is caused by the combined action of the spin-orbit
coupling, absence of inversion symmetry, and the time-reversal
symmetry-breaking in the electric field\cite{Edel, Inoue}. Since the
in-plane spin accumulation has been investigated in detail in previous
literatures and it has no relation with the intrinsic spin Hall effect, we
will not discuss it again in the present paper. Unlike the in-plane spin
accumulation, the perpendicular spin accumulation given by Eq.(38) is caused
by the intrinsic spin\ Hall effect, so its spatial distribution is highly
inhomogeneous and its magnitude depends sensitively on the spin diffusion
length $D$ and on the sample width $w$. This was illustrated clearly in
Fig.2, where we have plotted the transverse spatial distributions of the
perpendicular spin accumulation in three distinct cases with different
ratios of $D/w$. From Fig.2 and Eq.(38), one can see that the perpendicular
spin accumulation is maximum at the edges of the sample, and the
perpendicular spin accumulation at the edges of the sample will increase
with the increase of the width of the sample. When the width $w$ of the
sample is much larger than the spin diffusion length $D$, the perpendicular
spin accumulation at the edges of the sample will approach a maximum value
of $\frac{eE_x}{4\pi }(\frac{m\tau _F}{\tau _F^{\uparrow \downarrow
}\epsilon _F})^{1/2}(1+\frac{\alpha ^2m}{\hbar ^2\epsilon _F})$, which is
independent of the sample width $w$. This will be a merit for the
experimental measurement of the effect. In order to get a quantitative
estimation of the perpendicular spin accumulation, let us consider some
actual experimental parameters. In current 2DEG high quality samples\cite%
{jdq, Others, jth}, the typical carrier concentrations range from $5\times
10^{11}\text{ to }10^{12}\text{ cm}^{-2}$, the strength of the Rashba
spin-orbit coupling is on the order of $1\times 10^{-11}\sim 5\times 10^{-11}%
\text{eVm}$, the effective mass of conduction electrons is about $0.05m_e$,
the relaxation time is typically $1$ps,$\text{ }$the spin diffusion length
is about $1\mu $m, and the Fermi energy $\epsilon _F$ is about $20\sim 50%
\text{meV.}$ If one consider a sample with the width $w=10\mu $m ( much
larger than the spin diffusion length ) and the external field $eE_x=10\text{%
KeV/m}$, then from Eq.(38) one can estimate that the perpendicular spin
accumulation at the edges of the sample can be as larger as $10^{-23}\text{J}%
\cdot \text{s}/m^2$. This magnitude should be large enough to be detected
experimentally.

\begin{figure}[htbp]
\includegraphics[width=0.35\textwidth]{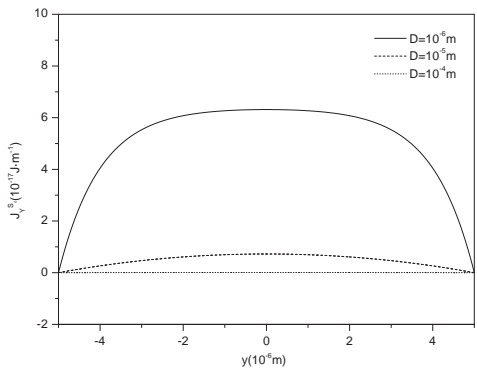}
\caption{Illustration of the transverse spatial distributions of
the spin Hall currents in three distinct cases with different
ratios of $D/w$. The parameters used are: the sample width
$w=10\protect\mu $m; the spin diffusion length $D=1\protect\mu $m
(the solid line), $10\protect\mu $m (the dashed line), and
$100\protect\mu $m\ (the dotted line).}
\end{figure}

\begin{figure}[tbp]
\includegraphics[width=0.35\textwidth]{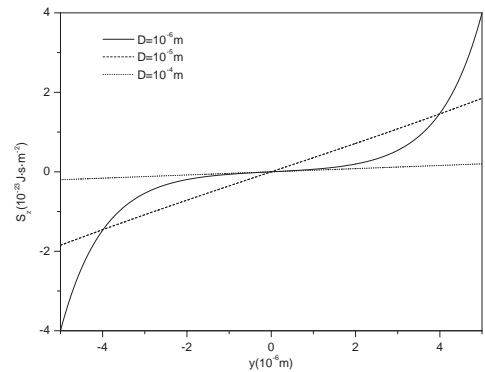}
\caption{The transverse spatial distributions of the perpendicular
spin accumulation in three distinct cases with different ratios of
$D/w$. ( The parameters used are: the sample width $w=10$
$\protect\mu $m; the spin diffusion length $D=1\protect\mu $m (the
solid line), $10\protect\mu $m (the
dashed line), and $100\protect\mu $m (the dotted line); the external field $%
eE_x=10\text{KeV/m; t}$he Rashba spin-orbit coupling constant $\protect%
\alpha =1\times 10^{-11}\text{eVm; }$the effective mass $m=$
$0.05m_e$; the
relaxation time is $1$ps; and the Fermi energy $\protect\epsilon _F=$ $20%
\text{meV.}$)}
\end{figure}

In conclusion, in this paper we have investigated the influences of spin
accumulation on the intrinsic spin Hall effect in two dimensional electron
gases with Rashba spin-orbit coupling. We have presented a detailed
theoretical analysis on the interplay between the spin Hall current and spin
accumulation in the intrinsic spin Hall effect in a Rashba two dimensional
electron gas. We have shown that in the presence of spin accumulation, the
spin Hall conductivity will not have a universal value. The spin Hall
current and spin accumulation in narrow strips of two dimensional electron
gases with Rashba spin-orbit coupling was calculated explicitly. The results
show that in order to calculate correctly the spin Hall current and the spin
Hall conductivity in a real sample with boundaries, the influences of spin
accumulation need to be taken into account. Recently, E. I. Rashba pointed
out that the Hamiltonian (1) implies that there exist nonvanishing
dissipationless spin currents even in the thermodynamic equilibrium state (
i.e., in the absence of the external electric field ).\cite{Rashba04} These
background spin currents are not associated with real spin transports but
spurious effects caused by the lacking of the time-reversal symmetry implied
in the Hamiltonian (1). Due to this fact, a procedure for eliminating the
spurious effects of these background spin currents should be devised in
calculating transport spin currents if the background currents contribute to
the calculation. But for the intrinsic spin Hall effect discussed in the
present paper, the background spin currents do not contribute to the
calculation of the spin Hall current due to the following reasons. First,
the spin Hall current is polarized in the direction perpendicular to the 2D
plane, while the background spin currents are polarized in the 2D plane.
Second, the spin Hall current is a dynamic response of the spins to the
external electric field and will vanish in the absence of the electric
field, but the background spin currents are independent of the electric
field. Due to these reasons, the background spin currents do not present in
the calculation of the spin Hall current and hence don't need to be
considered in the present paper.

This work is supported by a grant from the Research Grant Council of Hong
Kong under Contract No. HKU 7088/01P and a CRCG grant from the University of
Hong Kong. Ma and Tao would acknowledge the support by National Natural
Science Foundation of China and 973 Project with No. 2002CB613504.


\begin{thebibliography}{99}
\bibitem{note1} Similar to Eq.(13), Eq.(27) is valid only if both spin-orbit
split bands are occupied. \ In the presence of spin accumulation,
the condition for both spin-orbit split bands are occupied is
$\epsilon _{F}-e\mu
^{+}>-m\alpha ^{2}/2\hbar ^{2}$ ( the bottom of the lower band ) and $%
\epsilon _{F}-e\mu ^{-}>0$ ( the bottom of the upper band ).
According to the relation of carrier density and Fermi level in
2DES
$n=\frac{2}{\pi}[(\frac{m\alpha}{\hbar^2})^2+\frac{m\epsilon_F}{\hbar^2}]$,
we can get that the carrier density should larger than a
$\alpha$-dependent value
$n>\frac{2}{\pi}(\sqrt{3}+5/2)(\frac{m\alpha}{\hbar^2})^2$. For
real materials, the band-dependent shifts $\mu ^{+}$ and $\mu
^{-}$ of the Fermi level are usually very small, so the occurence
of spin accumulation will not change the situation that whether
the two spin-orbit split bands are both occupied or not.
\end{thebibliography}
\end{document}